\documentclass[journal]{IEEEtran}
\ifCLASSINFOpdf

\else
  
\fi

\usepackage{cite}
\usepackage{amsfonts}
\usepackage{amsmath}
\usepackage[pdftex]{graphicx}
\usepackage{amsmath,amssymb,amsfonts}
\usepackage{graphicx}
\usepackage{booktabs} %表格线条加粗所需
\usepackage{textcomp}
\usepackage{color}
\usepackage{float}
\usepackage{makecell}
\usepackage[ruled,linesnumbered,vlined]{algorithm2e}
\usepackage{geometry}

\usepackage{hyperref}
\hypersetup{
            colorlinks=true,
            linkcolor=blue,
            anchorcolor=blue,
            citecolor=blue,
            filecolor=blue,
            urlcolor=blue
            }
\makeatletter
\newcommand{\ccc}[1]{{\color{blue}\cite{#1}}}
\makeatother

\usepackage{amsmath}

\definecolor{b}{rgb}{0,0,0}

\begin{document}

\title{Two-Tier Task Offloading for Satellite-Assisted Marine Networks: A Hybrid Stackelberg-Bargaining Game Approach}

 \author{Zhen~Wang,~\IEEEmembership{Student Member,~IEEE,}
         Bin~Lin,~\IEEEmembership{Senior Member,~IEEE,}
         Qiang (John) Ye,~\IEEEmembership{Senior Member,~IEEE,}
         Haixia~Peng,~\IEEEmembership{Member,~IEEE}
%         and~Xiaoling~Han,~\IEEEmembership{Student Member,~IEEE}% <-this % stops a space
\thanks{Copyright (c) 20xx IEEE. Personal use of this material is permitted. However, permission to use this material for any other purposes must be obtained from the IEEE by sending a request to pubs-permissions@ieee.org.}
\thanks{\emph{(Corresponding author: Bin Lin.) }}
\thanks{Zhen Wang is with the Department of Information Science and Technology of Dalian Maritime University, Dalian, China, and Communication Engineering of Dalian Neusoft University of Information, Dalian, China. E-mail: wangzhen\underline{~}jsj@neusoft.edu.cn.}% <-this % stops a space
\thanks{Bin Lin and is with the Department of Information Science and Technology of Dalian Maritime University, Dalian, China. E-mail: binlin@dlmu.edu.cn.}% <-this % stops a space
\thanks{Qiang (John) Ye is with the Department of Electrical and Software Engineering, Schulich School of Engineering, University of Calgary, 2500 University Drive NW, Calgary, AB T2N 1N4. Email: qiang.ye@ucalgary.ca.}% <-this % stops a space
\thanks{Haixia Peng is with the School of Information and Communications Engineering, Xi'an Jiaotong University, Xi'an 710049, China. E-mail:  haixia.peng@xjtu.edu.cn.}% <-this % stops a space
\thanks{This manuscript has been accepted by IEEE Internet of Things Journal, DOI:10.1109/jiot.2024.3523527.}

 }

\maketitle

\begin{abstract}
The proliferation of maritime activities has spurred the emergence of numerous computation-intensive and delay-sensitive marine applications and services. Given the inherent rationality, selfish nature, and limited computational abilities of marine devices, devising effective strategies to incentivize their participation in task processing become a critical challenge.
In this paper, we investigate the satellite-assisted marine multi-access edge computing (MEC) and propose a two-tier task offloading scheme through a hybrid Stackelberg-Bargaining game approach to enhance offloading efficiency and maximize the utility of marine devices. 
Specifically, for the underwater acoustic communication, we consider the scenario where multiple autonomous underwater vehicles (AUVs), managed by maritime autonomous surface ships (MASSs), upload their collected data using non-orthogonal multiple access(NOMA) to optimize channel utilization. For the data transmission above the sea surface, we consider the scenario where a low earth orbit satellite (LEOS) functions as a space edge server to provide computing services, and MASS offloads workloads to LEOS through frequency division multiple access(FDMA) to prevent co-channel interference. 
we define the utility of AUVs, MASSs and LEOSs, and model the offloading process between AUVs and MASSs as a Stackelberg game, while representing the offloading interaction between MASSs and LEOSs as a Bargaining game. Additionally, we propose efficient algorithms to optimize AUV offloading strategies and MASS pricing strategies, while refining the bidding strategies for both MASSs and LEOSs. Simulation results demonstrate that the proposed algorithms significantly outperform benchmark schemes in achieving optimal solutions. 

\end{abstract}


\IEEEpeerreviewmaketitle

\section{Introduction}

\IEEEPARstart{W}{ith} the increasing scope of maritime activities, a wide array of marine wireless devices, such as underwater sensors, autonomous underwater vehicles (AUVs), vessels, buoys, and maritime autonomous surface ships (MASSs), has been deployed across the ocean. This proliferation has driven the rapid expansion of marine wireless networks, resulting in the generation of vast amounts of computation-intensive and delay-sensitive data\ccc{10499911},\ccc{9678008},\ccc{10449425}. For instance, in underwater exploration and surveillance, large volumes of hydrological data and monitoring images/videos must be collected and transmitted to designated equipment for processing. Similarly, during intelligent reconnaissance and maritime rescue operations, a large volume of collected real-time video footage requires prompt processing to enable timely decision-making. 
These applications have sparked stringent requirements on high-speed marine communication and powerful computing\ccc{10518032}. It is of utmost importance to establish an efficient marine communication and computing network for marine devices\ccc{10376220}. Nevertheless, existing marine communication and computing networks fall short in fulfilling the substantial demands of marine devices due to their low data rate and high latency\ccc{10173733},\ccc{9881915}.

Multi-access edge computing (MEC) has emerged as an efficient solution to improve computing efficiency and reduce decision-making latency for maritime terminals\ccc{9454395},\ccc{10373079}. Recently, various research endeavors have been made in proposing diverse innovative approaches to enhance the overall performance and efficiency of maritime MEC systems. 
In\ccc{9978925}, Hou \emph{et al.} designed a multi-tier underwater computing framework for autonomous underwater vehicles (AUVs), aiming at collecting and forwarding data sensed by underwater devices to surface-stations. In\ccc{9750912}, Wen \emph{et al.} proposed an intelligent oceanic data collection scheme for multi-automatic underwater gliders in maritime wireless communication networks. 
To further enhance data collection and offloading efficiency across diverse transmission segments, non-orthogonal multiple access (NOMA) has been adopted in maritime MEC networks, especially for acoustic data transmissions\ccc{10100681},\ccc{9882311}. 

Six generation (6G) wireless technologies stand poised to revolutionize the wireless landscape by integrating satellite networks with ground networks, ushering in a new era of seamless connectivity and advanced functionality\ccc{10700788}. The edge computing paradigm holds significant potential for applications and services in satellite-assisted networks, enabling support for computation-intensive and delay-sensitive services in oceanic environments\ccc{9662985},\ccc{9839554}. 
Zeng \emph{et al.} proposed an architecture for space-air-sea integrated networks aiming at enhancing the security of maritime communication networks for unmanned surface vehicles in\ccc{10463010}. Wang \emph{et al.} proposed a space-air-sea integrated network architecture designed to provide secure MEC services for delay-sensitive applications for maritime mobile users\ccc{10092853}.
Given the limited energy and computing resources of offshore equipment, integrating satellite and maritime networks with MEC presents a promising solution for enhancing the computing efficiency of marine applications.

However, the features of a marine environment pose technical challenges in fulfilling various oceanic application requirements\ccc{10490255}. 
First, as an integral component of marine networks, underwater acoustic communication is typically affected by unfavorable factors, such as limited data rates, prolonged propagation delay, and elevated packet loss rates\ccc{10118952},\ccc{9779873}. It impedes the on-demand service provisioning and limits the potential of maritime applications in areas such as monitoring, exploration, and emergency response\ccc{9741462},\ccc{10220230}. 
Second, unlike terrestrial networks, oceanic communication and computing resources are extremely limited. Current standalone oceanic network infrastructures are unable to meet the demands of numerous computation-intensive and delay-sensitive tasks at sea. Third, due to the heterogeneity of satellite-marine integrated networks, different network service providers may lack sufficient incentives to offer edge computing services for different oceanic applications\ccc{10118952}.
It is imperative to design an efficient maritime communication and computing scheme to overcome the aforementioned challenges and better serve the evolving demands of maritime applications and services.

Currently, how to optimize the computation offloading for AUVs with respect to the space-sea integrated network architecture is still an open issue. Motivated by the above observations, in this paper, we propose a satellite-assisted maritime MEC architecture to enhance task offloading efficiency and optimize the utilization of marine wireless devices. Also, a two-tier task offloading scheme is exploited for AUVs through a hybrid Stackelberg-Bargaining game approach. 
The key contributions of this paper mainly include the following aspects:
\textcolor{b}{
\begin{itemize}
\item[$\bullet$] 
\emph{Two-Tier Satellite-Assisted Maritime Task Offloading}: We first propose a low earth orbit satellite (LEOS) assisted task offloading framework in maritime MEC, which consists of two transmission tiers, i.e., the underwater acoustic communication and the satellite-assisted wireless communication. For underwater acoustic communication, multiple AUVs collect and upload oceanic data to MASSs utilizing NOMA to improve the underwater acoustic channel utilization. For satellite-assisted communication, MASSs transmit partial workloads to LEOSs for further processing through FDMA to avoid co-channel interference. 
\item[$\bullet$] 
\emph{Utility Maximization with Incentive-Driven Approaches}: We employ an incentive-based framework to encourage marine wireless devices (i.e., AUVs, MASSs and LEOSs) to participate in task computing activities, with the goal of optimizing their respective profits. 
Specifically, for underwater acoustic communication, we model the offloading process as a Stackelberg game to optimize AUV offloading strategies and MASS pricing strategies. For
MASS-to-LEOS transmission, we model the offloading process as a Bargaining game to refine the bidding strategies of MASSs and LEOSs, ensuring their mutual benefits. By analyzing the interactions among AUVs, MASSs, and LEOSs, we identify the game equilibrium that maximizes system efficiency and profitability.
\item[$\bullet$] 
\emph{Performance Evaluation}: We perform extensive simulations to validate the efficacy of the proposed offloading and incentive schemes. Numerical results indicate that the proposed algorithms optimize participant utilities, demonstrating their effectiveness and efficiency compared to benchmark schemes.
\end{itemize}
}
The remainder of this paper is organized as follows. Section II provides an overview of related works. The system model is presented in Section III. The problem formulation and solution design are given in Section IV. Section V presents the performance evaluation, and Section VI presents concluding remarks and future work. 
\textcolor{b}{Table I outlines the key notations in the sequence of their initial appearance in the paper.}

 \begin{table*}[htbp]\color{b}
 \centering
 \caption{Key Notations}
 \begin{tabular}{|c|l|c|l|}%l=left, r=right,c=center分别代表左对齐，右对齐和居中，字母的个数代表列数
 \hline
 $\mathbf{Notation}$ & $\mathbf{Description}$ &$\mathbf{Notation}$ & $\mathbf{Description}$  \\ 
 \hline
 $\Upsilon_m$  &Spatial coordinates of MASS $m$ & $\Upsilon_i$  & Spatial coordinates of AUV $i$  \\
 \hline
 $\lambda_i$  &Task generated by AUV $i$ &$D_i^{tot}$  & Input data size (in bits) of $\lambda_i$\\
 \hline
 $C_i^{tot}$  &Number of CPU cycles for processing $\lambda_i$&$a_{i,m}$  & Offloading ratio of AUV $i$ to MASS $m$\\
 \hline
 $b_{i,n}$  &Ratio of offloading task $\lambda_i$ from MASS $m$ to LEOS $n$ & $d_{i,m}$  &Distance between AUV $i$ and MASS $m$\\
\hline
$\varrho$  &Coefficient of $d_{i,m}$&$\Omega(f)$  & Absorption coefficient\\
\hline
$f$  &Central frequency of the acoustic signal&$\sigma$  &Spreading factor\\
\hline
$g_{i,m}$  &Acoustic uplink channel gain from AUV $i$ to MASS $m$ &
$N_B$  &Oceanic noise power spectrum density  \\
\hline
$W_i$  &Allocated channel bandwidth of AUV $i$ &
$R_{i,m}$  &Link transmission capacity from AUV $i$ to MASS $m$  \\
\hline
$p_i$  &Transmission power of AUV $i$  &
$s_i$  &Underwater link quality indicator \\
\hline
$t_{i,m}^{ts}$  &Transmission time of AUV $i$  &
$E_{i,m}^{ts}$  &Energy consumption of AUV $i$ for transmitting tasks  \\
\hline
$d_{m,n}$  &Distance between  MASS $m$ and LEOS $n$  &
$\varsigma_{m,n}$  &Geocentric angle between MASS $m$ and LEOS $n$  \\ 
\hline
$\theta_{m,n}$  &Elevation angle between MASS $m$ and LEOS $n$  &
$T^{\max}$  & Maximum communicate of each LEOS \\
\hline
$v_n$  &Moving speed of LEOS $n$   &
$R_{m,n}$  &Transmission rate of MASS $m$   \\
\hline
$\alpha_m$  &Bandwidth allocation factor &
$p_m$  &Transmission power of MASS $m$  \\
\hline
$g_m$  &Complex Gaussian variable &
$\beta_m$  &Fading involving shadowing, rain, and other fading \\
\hline
$\chi$  &Path exponent &
$t_{i,n}^{ts}$  &\makecell[l]{Communication latency for transmitting partial \\workloads of AUV $i$ from MASS $m$ to LEOS $n$}  \\
\hline
$E_{i,n}^{ts}$  &\makecell[l]{Energy consumption of MASS $m$ for transmitting \\partial workloads of AUV $i$ from MASS $m$ to LEOS $n$}&
$t_i^{com}$  &Computing time of AUV $i$ \\
\hline
$E_i^{com}$  &Energy consumption $E_i^{com}$ of AUV $i$ for computing&
$\rho_i/\rho_m/\rho_n$  &Computation capacity of AUV $i$ /MASS $m$ /LEOS $n$ \\
\hline
$\varepsilon_i/\varepsilon_m/\varepsilon_n$  &\makecell[l]{Effective power consumption coefficient of AUV $i$\\ /MASS $m$ /LEOS $n$}&
$t_{i,m}^{tc}$  &Transcoding time of MASS $m$\\
\hline
$E_{i,m}^{tc}$  &Energy consumption for transcoding data of MASS $m$&
$t_{i,m}^{com}$  &Latency of MASS $m$ to complete the assigned task \\
\hline
$E_{i,m}^{com}$  &\makecell[l]{Energy consumption of MASS $m$ for completing the\\ assigned task}&
$t_{i,n}^{com}$  &Latency of LEOS $n$ to complete the assigned task \\
\hline
$E_{i,n}^{com}$ &\makecell[l]{Energy consumption of LEOS $n$ for completing the\\ assigned task}&
$T_i^{tot}$  &Overall latency for completing AUV $i$’s workloads \\
\hline
$u_i$  &SoA function of AUV $i$ &
$\varpi_i$  &Initial satisfaction of AUV $i$ \\
\hline
$\xi_i$  &Calibration factor of the SoA function&
$Z_i/Z_m/Z_n$  &Utility function of AUV $i$ /MASS $m$ /LEOS $n$ \\
\hline
$q_i/q_m/q_n$  &Energy cost of AUV $i$ /MASS $m$ /LEOS $n$ &
$k_i$  &Price for each bit of data paid by AUV $i$  \\
\hline
$u_{i,m}$  &Satisfaction functions of MASS $m$  &
$u_{i,n}$  &Satisfaction functions of LEOS $n$  \\ 
\hline
$\omega_{i,m}$  &Weighting parameter of MASS $m$&
$\omega_{i,n}$  &Weighting parameter of LEOS $n$ \\ 
\hline
$\phi_{i,m}(b_{i,n})$  &Tender price of MASS $m$ &
$\phi_{i,n}(b_{i,n})$  &Tender price of LEOS $n$ \\ 
\hline
$\psi_{i,m}^{price}(b_{i,n})$  &Final transaction price&
$\eta_i$  &Profit distribution ratio \\ 
\hline
$\mu_{i,m}$  &Discount factor of MASS $m$  &
$\mu_{i,n}$  &Discount factors LEOS $n$\\
\hline
$\gamma_{i,m}$  &Patience coefficient of MASS $m$ &
$\gamma_{i,n}$  &Patience coefficient LEOS $n$\\

 \hline
 \end{tabular}
 \end{table*}

\section{Related work}
In this section, we provide an overview of existing works in multi-tier task offloading schemes, satellite-assisted maritime networks, and game theory based MEC, respectively.
\textcolor{b}{For brevity, comparisons with representative existing works are summarized in Table II.}
\begin{table*}[h!]\color{b}

	\begin{center}
		\caption{Related Work.}
		\resizebox{\textwidth}{!}{
		\begin{tabular}{|c|l|l|l|} % <-- Alignments: 1st column left, 2nd middle and 3rd right, with vertical lines in between
			\hline
			\textbf{References} & \textbf{Network type} & \textbf{Offloading} & \textbf{Comparisons with existing studies}\\		
			\hline
			\ccc{10449425} & {\makecell[l]{Air-sea integrated\\ networks}} & UAVs$\to$USV fleets & {\makecell[l]{The consideration is limited solely to the airspace segment, disregarding\\ the underwater segment and space segment.}}\\
			\hline
			\ccc{10518032} & {\makecell[l]{Space-air-ground-sea\\ integrated networks}} & {\makecell[l]{M-IoT$\to$LAP$\to$LEOS;\\
			M-IoT$\to$BS$\to$buoys }}& {\makecell[l]{Acoustic communication for underwater marine devices is not taken into\\ consideration.}}\\
			\hline
			\ccc{10376220} & {\makecell[l]{Air-sea integrated\\ networks}} & M-IoT$\to$UAVs & {\makecell[l]{The consideration is limited solely to the airspace segment, disregarding\\ the underwater segment and space segment.}}\\
			\hline
			\ccc{10092853} & {\makecell[l]{Space-air-sea\\ integrated networks}} & {\makecell[l]{M-IoT$\to$BS\\M-IoT$\to$UAV$\to$LEOS}} & {\makecell[l]{Acoustic communication for underwater marine devices is not taken into\\ consideration. }}\\
			\hline
			\ccc{10490255} & {\makecell[l]{Space-air-sea\\ integrated networks}} & {\makecell[l]{Surveillance UAV\\$\to$relay UAV}} & {\makecell[l]{Acoustic communication for underwater marine devices is not taken into\\ consideration.}}\\
			\hline
			\ccc{9741462} & {\makecell[l]{Ground-sea \\integrated networks}} & Vessel$\to$BS/buoy & {\makecell[l]{The consideration is limited solely to the offshore area, disregarding the\\ underwater segment and space segment.}}\\
			\hline
			\ccc{9693477} & {\makecell[l]{Air-ground-sea\\ integrated networks}} & {\makecell[l]{Vessel$\to$BS\\ relayed by UAVs}}  & {\makecell[l]{The consideration is limited solely to the offshore area and airspace seg-\\ment, disregarding the underwater segment and space segment.}}\\
			\hline
			\ccc{10453323} & {\makecell[l]{Ground-sea\\ integrated networks}} & Vessel$\to$buoy$\to$BS & {\makecell[l]{The consideration is limited solely to the offshore area, disregarding the\\ underwater segment and space segment.}}\\
			\hline
			\ccc{10224351} & {\makecell[l]{Air-ground-sea\\ integrated networks}} & {\makecell[l]{USV$\to$BS\\ USV$\to$UAV}}  & {\makecell[l]{The consideration is limited solely to the offshore area and airspace seg-\\ment, disregarding the underwater segment and space segment.}}\\
			\hline
			\ccc{7932121} & {\makecell[l]{Underwater wireless\\ sensor networks}} & nodes$\to$AUV& {\makecell[l]{The consideration is limited solely to the underwater segment.}}\\
			\hline
			\ccc{10279604} & {\makecell[l]{Space-air-ground-sea\\ integrated networks}} & M-IoT$\to$UAV$\to$vessel& {\makecell[l]{Acoustic communication for underwater marine devices is not taken into\\ consideration.}}\\
			\hline
			\ccc{9996964} & {\makecell[l]{Air-ground-sea\\ integrated networks}} & USV$\to$HAP$\to$BS  & {\makecell[l]{The consideration is limited solely to the offshore area and airspace seg-\\ment, disregarding the underwater segment and space segment.}}\\
			\hline
			\ccc{10254489} & {\makecell[l]{Space-air-ground-sea\\ integrated networks}} & No task offloading & {\makecell[l]{The focus is primarily on network performance optimization.}}\\
			\hline
			\ccc{9941481} & {\makecell[l]{Space-air-sea\\ integrated networks}} & No task offloading& {\makecell[l]{The focus is primarily on network performance optimization.}}\\
			\hline
			\ccc{9979795} & {\makecell[l]{Space-air-ground-sea\\ integrated networks}} &No task offloading& {\makecell[l]{The focus is primarily on network performance optimization.}}\\
			\hline
			\ccc{10414013} & {\makecell[l]{Space-air-ground-sea\\ integrated networks}} &No task offloading& {\makecell[l]{The focus is primarily on network performance optimization.}}\\
			\hline
			\ccc{10660307} & {\makecell[l]{Space-air-ground-sea\\ integrated networks}} &No task offloading& {\makecell[l]{The focus is primarily on network performance optimization.}}\\
			\hline
			Our work & {\makecell[l]{Space-sea-underwater\\ integrated networks}} & AUV$\to$MASS$\to$LEOS& {\makecell[l]{A satellite assisted maritime MEC network for AUVs is proposed and \\the acoustic communication is taken into consideration.}}\\
	        \hline
		
		\end{tabular}	}
	\end{center}

\end{table*}

\subsection{Multi-tier Task Offloading Schemes}
Multi-tier task offloading has emerged as an efficient solution to improve the computing efficiency of marine devices\ccc{10092853}. Many studies have explored various task offloading strategies in marine networks, which is classified into two categories. One approach is shore-based task offloading schemes, which necessitates land-based stations to offer computing services to offshore devices, typically with the aid of relay devices like unmanned aerial vehicles (UAVs) to transmit offshore computing tasks to shore-based stations\ccc{9741462},\ccc{9693477}. 
Wang \emph{et al.} established an MEC-enabled sea lane monitoring network architecture where the tasks generated by ships are offloaded to surface-stations and coastal base stations for processing in\ccc{10453323}. 
Dai \emph{et al.} proposed a hybrid offshore and aerial-based MEC scheme where both offshore base-stations and UAVs can provide computing services for marine unmanned surface vehicles in\ccc{10224351}. 
\textcolor{b}{Another type of approach is off-shore task offloading schemes, which involve unmanned devices (e.g., MASSs\ccc{10449425}, AUVs\ccc{7932121} and UAVs\ccc{10376220},\ccc{10279604}) and high-altitude devices\ccc{9996964} facilitating edge computing services for offshore devices. 
Zeng \emph{et al.} proposed an energy-efficient USV fleets-assisted computation offloading framework, where UAVs serve as the requesters for computing services, and USV fleets are the helpers in\ccc{10449425}. 
In\ccc{7932121}, Gjanci \emph{et al.} focused on underwater multi-modal wireless sensor networks (UWSNs) for submarine surveillance and monitoring applications, where nodes offload data to an AUV through optical technology and employ acoustic communication for coordination.
Lyu \emph{et al.} proposed a computation offloading and resource allocation scheme for NOMA-UAV-assisted maritime emergency communications to minimize computation overhead in\ccc{10376220}. 
You \emph{et al.}  proposed a computation offloading framework for uncertain marine tasks, where both UAVs and vessels provide computing services to marine Internet of Things (M-IoT) in\ccc{10279604}. 
Li \emph{et al.} proposed a secure computation offloading scheme for M-IoT with the assistance of unmanned surface vehicles (USVs) subject to the eavesdropping attack, where the high altitude platform (HAP) performs computation offloading for marine devices in\ccc{9996964}. }

\textcolor{b}{Different from existing offloading schemes, we propose a satellite assisted maritime MEC network for AUVs, to enable workload offloading to satellites for further computing, considering that satellites cover remote sea areas and have sufficient on-board energy capacity. We take the underwater acoustic communication into consideration and employ NOMA to optimize the underwater acoustic channel utilization. Additionally, we account for the underwater acoustic link quality when optimizing offloading decisions. }

\subsection{Satellite-assisted Marine Networks}
To address the ubiquitous communication and computing demands of marine devices, satellite-assisted maritime networks has recently surfaced as a novel research topic\ccc{10254489},\ccc{9941481}. 
%Satellite-terrestrial integrated networks related architectures, task offloading schemes, and challenges have been actively studied\ccc{10147866},\ccc{10198334},\ccc{10227270}.
%He \emph{et al.} proposed a load-balanced collaborative offloading strategy aiming at achieving the balanced traffic distribution in LEO satellite networks in\ccc{10144545}. Zhang \emph{et al.} formulated a Multi-Hop Satellite Peer offloading problem to jointly minimize the delay and energy consumption for satellite networks\ccc{10107794}. 
Li \emph{et al.} proposed an energy-efficient scheme for the space-air-ground-sea-integrated network(SAGSIN), utilizing an intelligent autonomous underwater glider specifically designed for marine applications in\ccc{9979795}.
To provide excellent quality of service for marine devices, Zhang \emph{et al.} proposed a multi-domain virtual network embedding solution for SAGSIN in\ccc{10414013}, while Lin \emph{et al.} proposed a resource management scheme employing software-defined networking (SDN) controllers to provide services for marine application in SAGSIN in\ccc{10660307}. 
Simultaneously, individual scholars have delved into satellite-assisted maritime MEC. Lin \emph{et al.} and Wang \emph{et al.} leverage low altitude platforms (LAPs)\ccc{10518032} and UAVs\ccc{10092853}, respectively, to assist in computing tasks, whereas some researchers primarily leverage satellites for signaling or navigation of the target area\ccc{10376220},\ccc{10490255},\ccc{10279604}. 

Inspired by the aforementioned studies, it is highly demanding to introduce satellite-assisted MEC into marine networks to provide oceanic devices with computation services by bringing the extensive computational resources of cloud servers to LEOSs. 

\subsection{Game theory based MEC}
Game theory has been extensively applied in MEC networks to establish distributed mechanisms and circumvent the high complexity inherent in centralized frameworks\ccc{10376220},\ccc{10102098},\ccc{10689514}. Zhang \emph{et al.} formulated a Stackelberg game based three-tier multi-user multi-task offloading model by considering users as leaders and BSs and the cloud as followers in\ccc{10452851}. Niu \emph{et al.} formalized a potential game for multi-mobile devices and multi-base stations to optimize latency sensitive tasks and latency-tolerant tasksin\ccc{10436712}. Peng \emph{et al.} leveraged a Stackelberg
game to model the dynamic interaction between edge server providers and mobile users and proposed an alternating direction method of multipliers algorithm to derive the optimal strategie in\ccc{10400778}. The above studies have investigated the interaction between selfish game participants and achieved a mutually satisfactory optimization solution to maximize their profits.

\textcolor{b}{Motivated by the above incentive-based schemes, in this paper, considering the rationality and selfishness of marine wireless devices, we investigate the satellite assisted maritime MEC framework and propose a hybrid Stackelberg-bargaining game approach to maximize the utilities of game participants and improve the computing efficiency of marine networks.}

\section{System model}
This section presents the scenario for the proposed two-tier task offloading in marine MEC networks including the system model, communication model, computation model, and utility model as follows. 
\subsection{Network model}
We consider a space-sea integrated network which consists of $N$ LEOSs providing edge computing capabilities and robust communication coverage for MASSs, denoted by $\mathcal{N}=\left\{1, 2,..., n,..., N\right\}$, where $N$ denotes the total number of satellites within a constellation, as shown in Fig. 1. A group of AUVs, denoted by $\mathcal{I}=\left\{1, 2,..., i,..., I\right\}$, are distributed on the seabed to monitor the marine environment and collect oceanic data (e.g., marine biological images, aquaculture monitoring videos, and intelligence reconnaissance videos). A set of MASSs equipped with computing capacities, denoted by $\mathcal{M}=\left\{1, 2,..., m,..., M\right\}$ are dispatched on the sea surface to receive the oceanic data collected by the AUVs. 
We consider four stages of uplink workload transmission and processing in the satellite-sea integrated system: in Stage I, AUVs execute local computing and upload a portion of their workloads to the corresponding MASS that they are associated with for further computation via NOMA; In Stage II, each MASS receives the offloaded data from its connected AUVs via underwater acoustic communication and then transcodes the data for further processing; In Stage III, the MASS processes partial workloads locally and transmits the remaining workloads to LEOS for further computation through FDMA; In Stage IV, each LEOS receives data offloaded from its connected MASSs through wireless transmission links and performs computation. 
%LEOS equipped with edge servers possesses impressive computing capabilities and can significantly reduce the computing latency. Nevertheless, the transfer of MASS's workload to LEOS would inevitably lead to significant delays in transmission, resulting in considerable latency.

To represent the locations of each AUV and MASS, a three-dimension (3D) Cartesian coordinate system is employed. 
Denote $\Upsilon_m=(x_m,y_m,z_m)$ as the spatial coordinates of MASS $m$, where $x_m$, $y_m$, and $z_m$ represent its the longitude, latitude, and height, respectively. Correspondingly, the coordinates of AUV $i$ is denoted as $\Upsilon_i=(x_i,y_i,z_i)$.
We denote each task generated by AUV $i$ as $\lambda_i=\left\{D_i^{tot},C_i^{tot}\right\}$, where $D_i^{tot}$ denotes the input data size (in bits) and $C_i^{tot}$ indicates the number of CPU cycles needed to process the task $\lambda_i$. We assume that all tasks can be executed on different network terminals. Specifically, we denote variable $a_{i,m}\in [0, 1]$ as the offloading ratio of AUV $i$ to MASS $m$. Then, the processing workload of AUV $i$ is expressed as $(1-a_{i,m})D_i^{tot}$. Denote $b_{i,n}\in [0, 1]$ as the ratio of offloading task $\lambda_i$ from MASS $m$ to LEOS $n$, then the workload for AUV $i$ processed on MASS $m$ is expressed as $(1-b_{i,n})a_{i,m}D_i^{tot}$.
% It is noteworthy that we assume the MASS operates in full-duplex mode, enabling concurrent uploading and downloading processes. Furthermore, given the significantly reduced size of the processed data compared to the original data, the result can be sent back within one time slot and the corresponding latency can be omitted.
\begin{figure}[htbp]
\centerline{\includegraphics[width=0.45\textwidth]{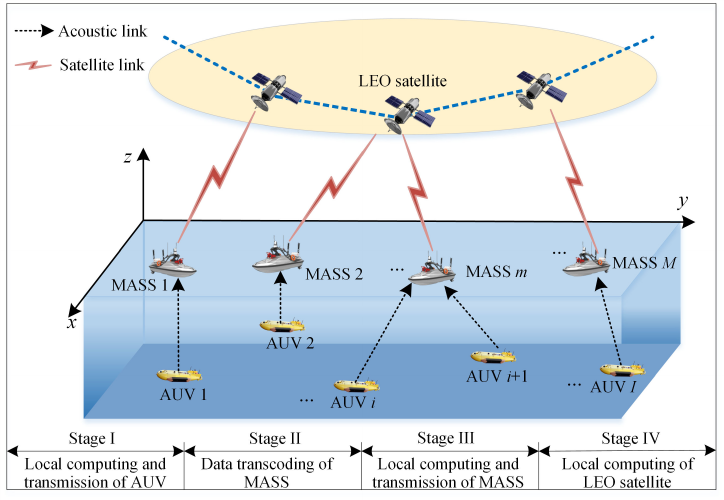}}
\caption{The satellite-sea integrated network architecture.}
\label{fig}
\end{figure}

\subsection{Communication model}

\emph{(1) Underwater acoustic communication from AUVs to MASSs}

Based on the Urick’s model\ccc{001}\ccc{4531073}, the transmission signal attenuation model tailored specifically for underwater acoustic communication between AUV $i$ and MASS $m$ is derived as:
\begin{equation}
H(d_{i,m},f)=d_{i,m}^\sigma\Omega(f)^{\varrho d_{i,m}},
\end{equation}
where $d_{i,m}$ is the distance between AUV $i$ and MASS $m$, $\varrho$ is the coefficient of $d_{i,m}$,  $\Omega(f)$ is the absorption coefficient, $f$ denotes the central frequency of the acoustic signal, and $\sigma$ is a spreading factor. The absorption coefficient $\Omega(f)$ can be obtained according to the Thorp's empirical formula in\ccc{4531073}.
The distance between AUV $i$ and MASS $m$ is expressed as
\begin{equation}
d_{i,m}=\sqrt{(x_m-x_i)^2+(y_m-y_i)^2+(z_m-z_i)^2}.
\end{equation}
Based on (1), the underwater acoustic uplink channel gain from AUV $i$ to MASS $m$ is calculated as
\begin{equation}
g_{i,m}=\frac{1}{H(d_{i,m},f)N_BW_i}.
\end{equation}
where $N_B$ is the oceanic noise power spectrum density and $W_i$ denotes the allocated channel bandwidth for acoustic communications from AUV $i$ to MASS $m$. Utilizing NOMA, the successive interference cancellation (SIC) necessitates an ordering based on their channel power gains relative to MASS $m$. The uplink channel gains at MASS $m$ from different UAVs using NOMA are sorted in a descending order as $g_{1,m}>g_{2,m}>...>g_{i,m}>...>g_{I,m}$.
According to Shannon theorem, the link transmission capacity from AUV $i$ to MASS $m$ can be calculated by
\begin{equation}
R_{i,m}=W_i\log_2\left(1+\frac{p_ig_is_i}{\sum_{j\in I,j\neq i}p_jg_js_j+N_B}\right), 
\end{equation}
where $p_i$ is the transmission power of AUV $i$ and $s_i$ represents \textcolor{b}{link quality indicator} during the offloading process of AUV $i$, which will be introduced in Section III-D.
Then, the time to transmit a portion of task $\lambda_i$ from AUV $i$ to MASS $m$ is calculated as
\begin{equation}
t_{i,m}^{ts}=\frac{a_{i,m}D_i^{tot}}{R_{i,m}}.
\end{equation}
The corresponding energy consumption of AUV $i$ is expressed as
\begin{equation}
E_{i,m}^{ts}=p_it_{i,m}^{ts}=\frac{p_ia_{i,m}D_i^{tot}}{R_{i,m}}.
\end{equation}

\emph{(2) LOS Transmission from MASSs to LEOSs}

\emph{(i) Coverage Model of LEOSs:} 
Different from terrestrial networks, LEOSs location dynamically vary, affecting the communication between MASSs and LEOSs. According to\ccc{9100240}, the geometric relation between an LEOS and an MASS is depicted in Fig. 2, where $h$ denotes the the altitude of the LEOS orbit, $R_e$ is the radius of the earth and $d_{m,n}$ represents the distance between  MASS $m$ and LEOS $n$, which is calculated as
\begin{equation}
d_{m,n}=\sqrt{R_e^2+(R_e+h)^2-2R_e(R_e+h)\cos\varsigma_{m,n}}.
\end{equation}
In (7), $\varsigma_{m,n}$ is the geocentric angle between MASS $m$ and LEOS $n$, expressed as
\begin{equation}
\varsigma_{m,n}=\arccos\left(\frac{R_e}{R_e+h}\cdot \cos\theta_{m,n}\right)-\theta_{m,n},
\end{equation}
where $\theta_{m,n}$ is the elevation angle between MASS $m$ and LEOS $n$, which is obtained by
\begin{equation}
\theta_{m,n}=\arccos\left(\frac{R_e+h}{d_{m,n}}\cdot \sin\varsigma_{m,n}\right).
\end{equation}
Subsequently, we can determine the maximum communicate time between MASS $m$ and LEOS $n$, denoted as
\begin{equation}
T^{\max}=\frac{2(R_e+h)\cdot \varsigma_{m,n}}{v_n}
\end{equation}
where $v_n=\sqrt{\frac{K_0}{R_e+h}}$ is the moving speed of LEOS $n$ and $K_0$ is the Kepler constant.
Given the high-speed movement of LEOS, the communication time between an MASS and an LEOS is constrained by the LEOS coverage duration.
\begin{figure}[htbp]
\centerline{\includegraphics[width=0.4\textwidth]{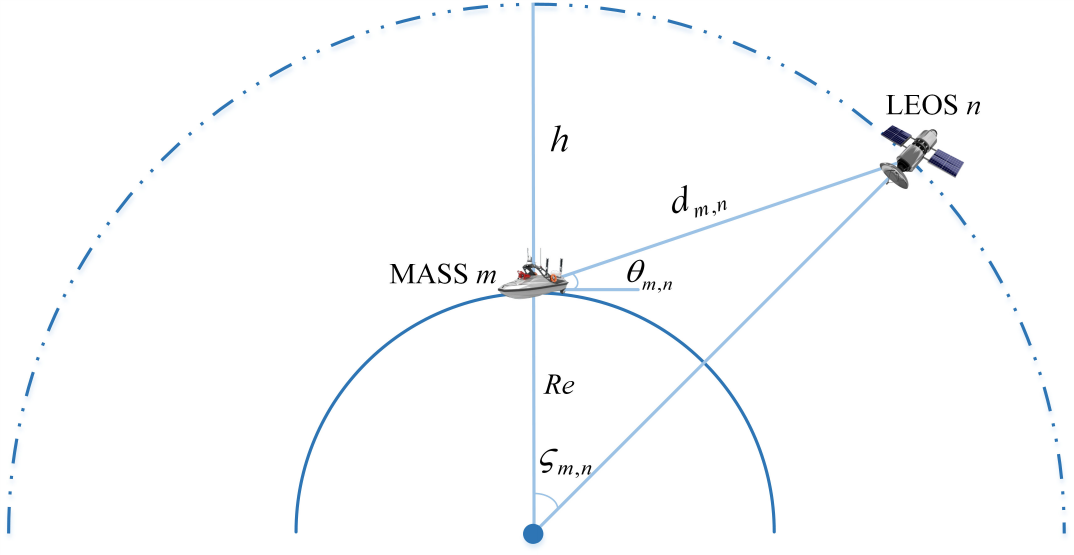}}
\caption{The geometric relation between LEOS $n$ and MASS $m$ in space.}
\label{fig}
\end{figure}

\emph{(ii) Communication Model from MASSs to LEOSs:}

The space segment comprises an LEO constellation, where each satellite is equipped with an MEC server to offer computing services to devices within its coverage area. Considering that all the MASSs share the spectrum resources of bandwidth $W$ via FDMA, the transmission rate of MASS $m$ is expressed as
\begin{equation}
R_{m,n}=\alpha_mW\log_2\left(1+\frac{p_m\mid h_{m,n} \mid^2}{\alpha_mWN_0}\right).
\end{equation}
In (11), $\alpha_m$ represents the proportion of bandwidth allocated to MASS $m$, $p_m$ denotes the transmission power of MASS $m$ and 
$h_{m,n}=g_m\cdot\beta_m\cdot(d_{m,n})^{-\chi}$, where $g_m$ is a complex Gaussian variable representing Rayleigh fading, $\beta_m$ denotes the fading involving shadowing, rain, and other fading, and $\chi$ is the path exponent.

However, the distance between an MASS and an LEOS is relatively long, making the propagation delay non-negligible. Thus, the communication latency for transmitting partial workloads of AUV $i$ between MASS $m$ and LEOS $n$ is calculated as
\begin{equation}
\textcolor{b}{t_{i,n}^{ts}}=\frac{a_{i,m}b_{i,n}D_i^{tot}}{R_{m,n}}+\frac{2d_{m,n}}{c},
\end{equation}
where $c$ denotes the speed of light.
The corresponding energy consumption of MASS $m$ is expressed as
\begin{equation}
\textcolor{b}{E_{i,n}^{ts}=p_mt_{i,n}^{ts}}=p_m\left(\frac{a_{i,m}b_{i,n}D_i^{tot}}{R_{m,n}}+\frac{2d_{m,n}}{c}\right).
\end{equation}

\subsection{Computation Model}

\emph{(1) Local computing of AUVs: }

Regarding local computation, we denote $\rho_i$ as the computation capacity (CPU cycles per second) of AUV $i$. The execution time of task $\lambda_i$ is given by 
\begin{equation}
t_i^{com}=\frac{(1-a_{i,m})C_i^{tot}}{\rho_i}.
\end{equation}
The corresponding energy consumption $E_i^{com}$ for AUV $i$ is computed as\ccc{10453323}:
\begin{equation}
E_i^{com}=\varepsilon_i\rho_i^3t_i^{com}=\varepsilon_i\rho_i^2(1-a_{i,m})C_i^{tot},
\end{equation}
where $\varepsilon_i$ is the effective power consumption coefficient of AUV $i$ in terms of computation. 

\emph{(2) Data transcoding of MASSs: }

MASS $m$ first receives $a_{i,m}D_i^{tot}$ of task $\lambda_i$ from AUV $i$ through acoustic communication and then transcodes the data for RF transmission. 
The transcoding time is
\begin{equation}
t_{i,m}^{tc}=\frac{\delta_ma_{i,m}D_i^{tot}}{\rho_m} 
\end{equation}
where $\delta_m$ denotes the number of CPU cycles for processing one information bit of acoustic signal and $\rho_m$ is the processing capacity of MASS $m$ in CPU cycles per second.
The energy consumption for transcoding data is expressed as
\begin{equation}
E_{i,m}^{tc}=\varepsilon_m\rho_m^3t_{i,m}^{tc}=\varepsilon_m\rho_m^2\delta_ma_{i,m}D_i^{tot}.
\end{equation}

\emph{(3) Local computing of MASSs: }

The latency of MASS $m$ to complete the assigned task is denoted as
\begin{equation}
t_{i,m}^{com}=\frac{(1-b_{i,n})a_{i,m}C_i^{tot}}{\rho_m}.
\end{equation}
The energy consumption for MASS $m$ in terms of task computation is calculated as
\begin{equation}
E_{i,m}^{com}=\varepsilon_m\rho_m^3t_{i,m}^{com}=\varepsilon_m\rho_m^2(1-b_{i,n})a_{i,m}C_i^{tot},
\end{equation}
where $\varepsilon_m$ is the effective power consumption coefficient of MASS $m$.

\emph{(4) Local computing of LEOSs: }

The partial computation tasks of AUV $i$ can be executed on the LEOS for further processing. We set $\rho_n$ as the computation capacity (CPU cycles/s) of LEOS $n$ and the execution time is calculated as
\begin{equation}
t_{i,n}^{com}=\frac{a_{i,m}b_{i,n}C_i^{tot}}{\rho_n}.
\end{equation}
Similar to MASS, the energy consumption of LEOS $n$ is calculated as
\begin{equation}
E_{i,n}^{com}=\varepsilon_n\rho_n^3t_{i,n}^{com}=\varepsilon_n\rho_n^2a_{i,m}b_{i,n}C_i^{tot},
\end{equation}
where $\varepsilon_n$ is the effective power consumption coefficient of LEOS $n$ in terms of task computation.

The overall latency for completing AUV $i$’s workloads is calculated as
\begin{equation}
T_i^{tot}=\max\left\{t_i^{com},t_{i,m}^{ts}+t_{i,m}^{tc}+\max\left\{t_{i,m}^{com},\textcolor{b}{t_{i,n}^{ts}}+t_{i,n}^{com}\right\}\right\}.
\end{equation}

\subsection{Utility Model}

\emph{(1) Utility function of AUVs: }

To express the level of satisfaction of AUVs (SoA) in performing computation offloading, we introduce a logarithmic function to reflect the computation offloading intention of AUVs\ccc{10102274}. The SoA function is concave and provides a quantitative assessment of the AUV's satisfaction with offloading activities based on task offloading ratio $a_{i,m}$ to MASS $m$, given by
\begin{equation}
u_i=\text{ln}(1+s_ia_{i,m})+\varpi_i^2
\end{equation}
where $s_i$ represents the link quality indicator during the offloading process of AUV $i$, and $\varpi_i^2$$(\varpi_i\in[0,1])$ denotes the initial satisfaction of AUV $i$ before the establishment of the MASS-assisted MEC system\ccc{10102274}. The value of $\varpi_i^2$ is contingent upon the computing and communication capabilities of the neighboring MASS before the deployment of MEC system, as well as the AUV $i$’s individual computing requirements.
The performance analysis of the satisfaction function is depicted in Fig. 3. When $\varpi_i$ remains constant, AUV $i$'s satisfaction increases with the increase of $s_i$. When both $\varpi_i$ and $s_i$ are high, the satisfaction can reach its maximum even without offloading all tasks, as indicated by the blue line. When $s_i$ is low (shown as the light blue line), the satisfaction rate increases slowly with the offloading rate, which means that if the signal quality is poor, AUV $i$ may struggle to achieve high satisfaction even with substantial task offloading.
\begin{figure}[htbp]
\centerline{\includegraphics[width=0.4\textwidth]{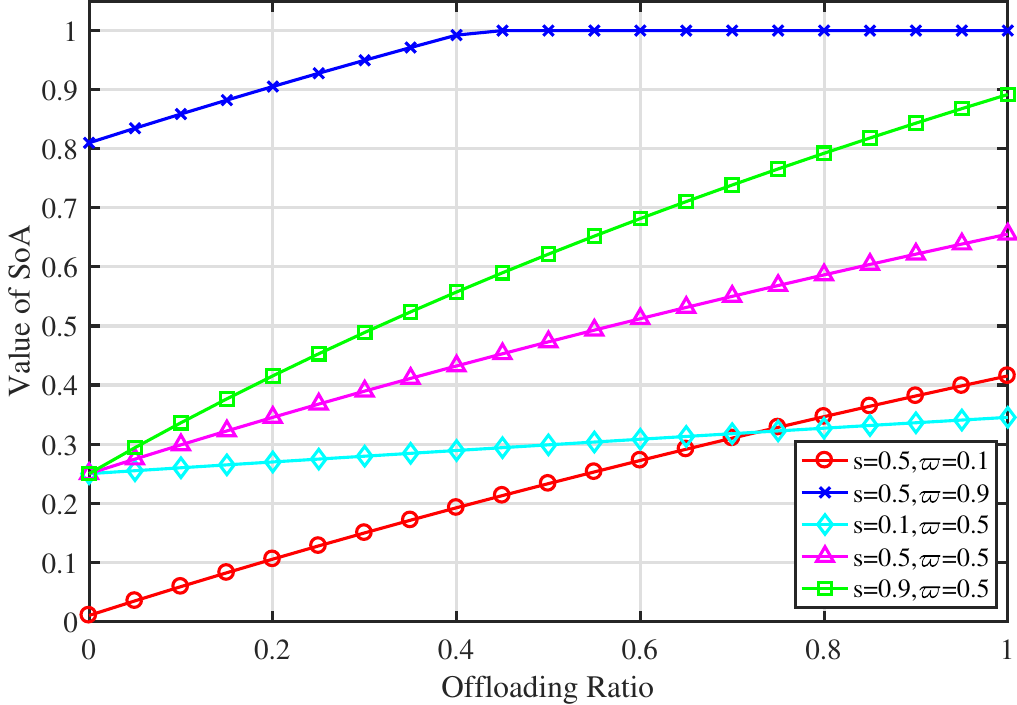}}
\caption{Performance analysis of $u_i$.}
\label{fig}
\end{figure}
The utility function of AUV $i$ is expressed as the difference between the SoA function of $u_i$, the price paid to MASS $m$ and the cost of completing its workload, given by
\begin{equation}
Z_i=\xi_i[\text{ln}(1+s_ia_{i,m})+\varpi_i^2]-k_ia_{i,m}D_i^{tot}-q_i(E_{i,m}^{ts}+E_i^{com}),
\end{equation}
where $\xi_i$ is a calibration factor of the SoA function to minimize the numerical discrepancy between $u_i$ and the associated workload processing cost, 
$k_i$ is the price for each bit of data that needs to be paid to MASS $m$ and $q_i$ is the cost of per unit of energy consumption.

\emph{(2) Utility function of MASSs: }

An MASS can assist in processing partial computation loads and its strategy is to determine the price for obtaining rewards from AUVs and the price paid to LEOS for assisting in computation workloads. For MASS $m$, the utility function is expressed as the difference between the revenue provide by AUV $i$, the price paid to LEOS, and the cost of processing its workloads, given by
\begin{equation}
Z_m=k_ia_{i,m}D_i^{tot}-\textcolor{b}{\psi_{i,m}^{price}(b_{i,n})}-q_m(\textcolor{b}{E_{i,n}^{ts}}+E_{i,m}^{tc}+E_{i,m}^{com}),
\end{equation}
where $q_m$ is the cost of per unit of energy consumption of MASS and \textcolor{b}{$\psi_{i,m}^{price}(b_{i,n})$} denotes the price paid to LEOS $n$ given the offloading ratio $b_{i,n}$, which will be introduced in Section IV-B.

\emph{(2) Utility function of LEOSs: }

An LEOS can assist in workload computation to earn rewards from MASSs. For LEOS $n$, the utility function is expressed as the difference between the total revenue provide by MASS $m$ and the cost of completing its workloads, given by
\begin{equation}
Z_n=\textcolor{b}{\psi_{i,m}^{price}(b_{i,n})}-q_nE_{i,n}^{com},
\end{equation}
where $q_n$ denotes the cost of per unit of energy consumption of LEOS $n$.

\section{Problem formulation and Solution Design}
This section presents our problem formulation and the solution design devised to tackle the aforementioned challenges. We first describe the offloading strategies of AUVs and the pricing strategies of each MASS as a Stackelberg game. We then establish a Bargaining game model to capture the interaction between an MASS and an LEOS and determine their bidding strategies. 
Based on the aforementioned models in section-III, the following optimization problems are formulated to maximize the utilities of AUV $i$, MASS $m$, and LEOS $n$, respectively.
For AUV $i$, the utility maximization problem is expressed as
\begin{align}
&\enspace\enspace\enspace(\textbf{P1}):\max_{a_{i,m}, k_i}Z_i \nonumber\\
&s.t.\enspace{\rm(C1)-\rm(C5)}.
\end{align} 
Similarly, the optimization problem of MASS $m$ is formulated as:
\begin{align}
&\enspace\enspace\enspace(\textbf{P2}):\max_{k_i, b_{i,n}}Z_m\nonumber\\
&s.t.\enspace{\rm(C5)-\rm(C8)}.
\end{align} 
For LEOS $n$, the optimization problem is formulated as:
\begin{align}
&\enspace\enspace\enspace(\textbf{P3}):\max_{b_{i,n}}Z_n\nonumber\\
&s.t.\enspace{\rm(C6), \rm(C9),\rm(C10)}.
\end{align} 
Where
\begin{align}
&{\rm(C1)}:0 \leq a_{i,m}\leq 1, \forall i\in \mathcal{I},\forall m\in \mathcal{M}, \nonumber\\
&{\rm(C2)}:E_i^{com}+E_{i,m}^{ts}\leq E_i^{\max}, \forall i\in \mathcal{I},\nonumber\\
&{\rm(C3)}:T_i^{tot}\leq T^{\max}, \forall i\in \mathcal{I},\nonumber\\
&{\rm(C4)}:Z_i\geq 0, \forall i\in \mathcal{I},\forall m\in \mathcal{M},\nonumber\\
&{\rm(C5)}:0\leq k_i\leq k_i^{\max}, \forall i\in \mathcal{I},\nonumber\\
&{\rm(C6)}:0 \leq b_{i,n}\leq 1, \forall m\in \mathcal{M},\forall n\in \mathcal{N}, \nonumber\\
&{\rm(C7)}:E_{i,m}^{com}+\textcolor{b}{E_{i,n}^{ts}}+E_{i,m}^{tc}\leq E_{i,m}^{\max}, \forall m\in \mathcal{M},\nonumber\\
&{\rm(C8)}:Z_m\geq 0, \forall i,m,n,\nonumber\\
&{\rm(C9)}:E_{i,n}^{com}\leq E_{i,n}^{\max}, \forall m\in \mathcal{M},\nonumber\\
&{\rm(C10)}:Z_n\geq 0, \forall m,n.
\end{align} 
In $(30)$, constraint (C1) and constraint (C6) indicate that the offloading ratio is in between $0$ and $1$. Constraint (C2), (C7) and (C9) guarantee that the energy consumption of each party cannot exceed the maximum value. Constraint (C3) ensures that the task completion time of AUV $i$ cannot exceed the maximum LEOS coverage time. Constraint (C4), (C8) and (C10) guarantee the utility of each party be higher than or equal to $0$. Constraint (C5) indicates that the price determined by AUV $i$ cannot exceed the maximum value.

\textcolor{b}{Since the variables in (\textbf{P1}), (\textbf{P2}), and (\textbf{P3}) are closely coupled, solving the optimization problems is non-deterministic polynomial-time hard (NP-hard). The solution space for the formulated problem is immense, and as the number of AUVs escalates, the computational effort increases exponentially, which is similar to the Traveling Salesman Problem (TSP, a well-known NP-hard problem)\ccc{1983Computers}.  Finding the optimal solution through a centralized approach is not tractable in polynomial time. To overcome these challenges, we develop a hybrid Stackelberg-Bargaining game framework to optimize the task offloading strategies and resource pricing policies in a distributed way with low computation complexity. }

\subsection{Stackelberg Game Formulation between AUVs and MASSs}
We describe the interaction between an AUV and an MASS as a Stackelberg game model. An MASS, say MASS $m$, is the leading player responsible for establishing the pricing strategies for task processing and reward acquisition. Subsequently, an AUV, say AUV $i$, as the follower, determines its offloading ratio in response to the price established by MASS $m$, with the objective of optimizing its overall revenue.
% Our goal is to maximize the utility of AUV $i$ and MASS $m$.
Given the predefined pricing strategies of MASS $m$, our initial step is to analyze the offloading ratio of AUV $i$ to optimize its utility. In accordance with the objective function of the problem, we derive the first-order derivative of $Z_i$ with respect to $a_{i,m}$ as
\begin{equation}
\frac{\partial Z_i(a_{i,m})}{\partial a_{i,m}}=\frac{\xi_is_i}{1+s_ia_{i,m}}-k_iD_i^{tot}+q_i\varepsilon_i\rho_i^2C_i^{tot}-\frac{q_ip_iD_i^{tot}}{R_{i,m}}.
\end{equation}
Then, the second derivative of $Z_i$ with respect to $a_{i,m}$ is calculated as
\begin{equation}
\frac{\partial^2 Z_i(a_{i,m})}{\partial a_{i,m}^2}=-\frac{\xi_is_i^2}{(1+s_ia_{i,m})^2}.
\end{equation}
As the second derivative of the utility function is negative, $Z_i$ is identified as a strictly concave function with respect to $a_{i,m}$, and the first derivative of $Z_i$ decreases with $a_{i,m}$, which proves the existence of Stackelberg equilibrium.

% The Nash equilibrium's existence verifies the constant presence of a Nash equilibrium point in the AUVs' data offloading strategy for a designated pricing strategy of the MASS. It is feasible to discern an optimal pricing decision profile alongside its corresponding data offloading decision profile, thus maximizing the utility for both AUVs and MASS. This underscores the existence of a subgame perfect Nash equilibrium within the Stackelberg game. 
Then, we obtain
\begin{equation}
\lim_{a_{i,m}\to0} Z_i^{'}(a_{i,m})=\xi_is_i-A
\end{equation}
and
\begin{equation}
\lim_{a_{i,m}\to1} Z_i^{'}(a_{i,m})=\frac{\xi_is_i}{1+s_i}-A
\end{equation}
where $Z_i^{'}(a_{i,m})=\partial Z_i(a_{i,m})/\partial a_{i,m}$ and $A=k_iD_i^{tot}-q_i\varepsilon_i\rho_i^2C_i^{tot}+q_ip_iD_i^{tot}/R_{i,m}$ in (33) and (34). The maximum utility of AUV $i$ is expressed as
\begin{equation}
\max_{a_{i,m}}Z_i=
\begin{cases}
Z_i(0),& \rm{if}\,\ \text{$ \xi_is_i-A<0$} \\
Z_i(a_{i,m}^*),& \rm{if} \,\ \text{$\frac{\xi_is_i}{1+s_i}-A\leq0\leq\xi_is_i-A$} \\
Z_i(1),& \rm{if}\,\ \text{$\frac{\xi_is_i}{1+s_i}-A>0$}
\end{cases}
\end{equation}
where $a_{i,m}^*$ is the root of $Z_i^{'}(a_{i,m}^*)=0$. In Algorithm 1, we propose a binary search algorithm (BSA) to obtain the value of $a_{i,m}^*$.

\begin{algorithm}
\SetAlgoLined
\caption{BSA Algorithm.}\label{algorithm}
\KwIn{ Given the tolerable computation-error $\delta$;}
\KwOut{The optimal value $a_{i,m}^*$ and the corresponding value of $Z_i(a_{i,m}^*)$; }
\textbf{Initialization:} Set the lower bound as $a_{i,m}^{l}$, set the upper bound as $a_{i,m}^{h}$\;
\While{$\vert a_{i,m}^{h}-a_{i,m}^{l}\vert>\delta$}
{ Update the current value of $a_{i,m}^{cur}=\frac{1}{2}(a_{i,m}^{h}+a_{i,m}^{l})$\;
  Obtain the value of $Z_i^{'}(a_{i,m}^{cur})$\;
  \eIf{ $Z_i^{'}(a_{i,m}^{cur})<0$}
  {
  Update the upper bound of the search range as $a_{i,m}^{h}=a_{i,m}^{cur}$\;
  }{
  	\eIf{$Z_i^{'}(a_{i,m}^{cur})>0$}
  	{
  		Update the lower bound of the search range as $a_{i,m}^{l}=a_{i,m}^{cur}$\;
  	}
   {Set $a_{i,m}^*=a_{i,m}^{cur}$\;
  	Calculate the value of $Z_i(a_{i,m}^*)$\;}
  }
 		   
}
\end{algorithm}

After obtaining the optimal offloading strategy of AUV $i$, similarly,we analyze (\textbf{P2}) to derive the pricing strategies of MASS $m$. To find the optimal value of $k_i$, we propose a linear search algorithm (LSA) within the given interval $[0,k_i^{max}]$  to numerically obtain the optimal solution of MASS $m$, as shown in Algorithm 2.

\begin{algorithm}
	\SetAlgoLined
	\caption{LSA Algorithm.}\label{algorithm}
	\textbf{Initialization:}  $j=0$, the current best value $Z_m^{cur}$, the current best price $k_i^{cur}$, the step size $\pi$\;
	Set $k_i(j)=k_i^{\min}$\;
	\While{$k_i(j)<k_i^{\max}$}
	{ Invoke Algorithm 1 to obtain $a_{i,m}^*$\;
	  Calculate $Z_m(k_i(j))$  according to Eq. (26)\;
		\If{ $Z_m^{cur}<Z_m(k_i(j))$ }
		{Update $Z_m^{cur}\leftarrow Z_m(k_i(j))$\;
		Update $k_i^{cur}\leftarrow k_i(j)$\;}
	 $j\leftarrow j+1$\;	
	 Update $k_i(j)$ with $k_i(j+1)=k_i(j)+\pi$\;	
	}
	Obtain the optimal price $k_i^*=k_i^{cur}$, $Z_m^*=Z_m^{cur}$\;
    Return $k_i^*$ and $Z_m^*$\;
	
\end{algorithm}

\subsection{Bargaining Game Formulation between MASSs and LEOSs}
We model the interaction between an MASS and an LEOS as a Bargaining game, where MASS $m$ is the buyer intending to purchase computation services for processing workloads, while LEOS $n$ serves as the seller to get revenue by providing computation resources. The satisfaction functions of MASS $m$ and LEOS $n$ are expressed, respectively, as
\begin{equation}
u_{i,m}=\omega_{i,m}\text{ln}\left(1+\frac{1}{t_{i,n}^{com}}\right)
\end{equation}
and
\begin{equation}
u_{i,n}=\omega_{i,n}\text{ln}\left(1+b_{i,n}\right),
\end{equation}
where $\omega_{i,m}$ and $\omega_{i,n}$ are the weighting parameters indicating the degrees of satisfaction of MASS $m$ and LEOS $n$, respectively. On this basis, we define the tender price for MASS $m$ as 
\begin{equation}
\phi_{i,m}(b_{i,n})=\min \left\{B,u_{i,m} \right\},
\end{equation}
where $B=k_ia_{i,m}D_i^{tot}-q_m(E_{i,n}^{ts}+E_{i,m}^{tc}+E_{i,m}^{com})$.
Similarly, the tender price for LEOS $n$ is formulated as
\begin{equation}
\phi_{i,n}(b_{i,n})=\max \left\{q_nE_{i,n}^{com},u_{i,n} \right\}.
\end{equation}
For MASS $m$, its satisfaction degree is higher if LEOS $n$ completes its offloaded workloads within a shorter time. 
For LEOS $n$, its satisfaction degree increases with the offloaded workload volume, indicating higher revenue. 
The final transaction price is calculated as
\begin{equation}
\psi_{i,m}^{price}(b_{i,n})=\eta_i\phi_{i,m}(b_{i,n})+(1-\eta_i)\phi_{i,n}(b_{i,n})
\end{equation}
where the weighting parameter $\eta_i(0\leq\eta_i\leq1)$ determines the weights of profit distribution between MASS $m$ and LEOS $n$. The optimal transaction price is determined through a Bargaining game, which involves two bargainers negotiating the distribution of the Bid-ask Spread $\phi_{i,m}(b_{i,n})-\phi_{i,n}(b_{i,n})$. If $\phi_{i,m}(b_{i,n})<\phi_{i,n}(b_{i,n})$, the transaction is cancelled, as the bidding price of MASS $m$ is less than that of LEOS $n$. If $\phi_{i,m}(b_{i,n})=\phi_{i,n}(b_{i,n})$, the transaction price is determined as $\phi_{i,m}(b_{i,n})$ or $\phi_{i,n}(b_{i,n})$. The bargaining game exists only when the bidding strategy satisfies $\phi_{i,m}(b_{i,n})>\phi_{i,n}(b_{i,n})$. In this case, MASS $m$ intends to minimize its expenses, while LEOS $n$ aims for maximizing its profit earning. Basically, it is a zero-sum game on the total transaction surplus $\phi_{i,m}(b_{i,n})-\phi_{i,n}(b_{i,n})$. After the transaction, LEOS $n$ obtains the remaining $\psi_{i,m}^{price}(b_{i,n})-\phi_{i,n}(b_{i,n})$, and MASS $m$ obtains the remaining $\phi_{i,m}(b_{i,n})-\psi_{i,m}^{price}(b_{i,n})$. 

According to Rubinstein’s bargaining model\ccc{002}, in an infinite round of Bargaining game, we complete the bargaining process within the first round and obtain the distinct subgame  Nash equilibrium as
\begin{equation}
\eta_i^*=\frac{\mu_{i,n}(1-\mu_{i,m})}{1-\mu_{i,m}\mu_{i,n}},
\end{equation}
where $\mu_{i,m}\in[0,1]$ and $\mu_{i,n}\in[0,1]$ are the discount factors which represent the respective patience levels exhibited by MASS $m$ and LEOS $n$. A higher patience level corresponds to a higher discount factor value, indicating a greater willingness to negotiate, while a lower patience level corresponds to a lower discount factor value, reflecting more impatience\ccc{003}. 
Since MASS $m$ aims to complete its computation workloads efficiently, its discount factor increases as the completion time for workload computation decreases. Then, $\mu_{i,m}$ is formulated as
\begin{equation}
\mu_{i,m}=\gamma_{i,m}\frac{\left\vert\text{ln}(1+t_{i,n}^{com})-\text{ln}\left(1+\frac{1}{t_{i,n}^{com}}\right)\right\vert}{\text{ln}(1+t_{i,n}^{com})+\text{ln}\left(1+\frac{1}{t_{i,n}^{com}}\right)}.
\end{equation}

Similarly, LEOS $n$ also strives to reap greater benefits by efficiently completing workloads and its discount factor $\mu_{i,n}$ increases with the increase of the offloading ratio $b_{i,n}$, given by
\begin{equation}
\mu_{i,n}=1-\gamma_{i,n}\frac{\left\vert\text{ln}(1+b_{i,n})-\text{ln}\left(1+\frac{1}{b_{i,n}}\right)\right\vert}{\text{ln}(1+b_{i,n})+\text{ln}\left(1+\frac{1}{b_{i,n}}\right)}.
\end{equation}
$\gamma_{i,m}$ and $\gamma_{i,n}$ indicate the the patience coefficients of MASS $m$ and LEOS $n$, respectively. Obviously, both players aim to reach a consensus on the proposed scheme for the transaction surplus distribution, as their utilities are diminished over time.

Based on the hybrid Stackelberg game and Bargaining game, we propose a two-tier task offloading algorithm (TTA) based on pricing-based incentive mechanisms, as described in Algorithm 3. The algorithm consists of two stages: the first stage (from step 2 to step 5) is to obtain the Stackelberg equilibrium of (\textbf{P1}) and (\textbf{P2}), and the second stage (from step 8 to step 15) to obtain the bargaining equilibrium of (\textbf{P2}) and (\textbf{P3}).
\begin{algorithm}
	\SetAlgoLined
	\caption{TTA Algorithm.}\label{algorithm}
    \KwIn{Given the tolerable computation-error $\delta$, the step size $\pi$;}
    \KwOut{$Z_m^*$ and $Z_n^*$;}
	\emph{Stage 1: Stackelberg equilibrium of} (\textbf{P1}) \emph{and} (\textbf{P2})\;
	\For{$k_i\in[0,k_i^{max}]$}{
	  Invoke Algorithm 1 to calculate the value of $a_{i,m}^*$\;
	  Invoke Algorithm 2 to obtain the optimal price $k_i^*$\;
	}
	\textbf{Input:} the Stackelberg equilibrium as $\left\{a_{i,m}^*,k_i^*\right\}$\;
	\emph{Stage 2: Bargaining equilibrium of} (\textbf{P2}) \emph{and} (\textbf{P3})\;	
	 Calculate the offloading workload $a_{i,m}^*D_i^{tot}$\;	
	 \eIf{$a_{i,m}^*\in(0,1]$}{
	    Calculate $\eta_i^*$ with Eq.(41)\;
	     Calculate $\phi_{i,m}(b_{i,n})$ and $\phi_{i,n}(b_{i,n})$ with Eq.(38) and Eq.(39)\;
	     Calculate  the optimal bidding price $\psi_{i,m}^{price}(b_{i,n})$ with Eq.(40)\;
	 }
	{(\textbf{P2}) and (\textbf{P3}) are infeasible\;}
	
\end{algorithm}

\textcolor{b}{We analyze the computational complexity of the proposed algorithms.
The complexities of Algorithm 1 and Algorithm 2 largely depend on the number of iterations required for the algorithms to converge. We denote the number of iterations of Algorithm 1 and Algorithm 2 as $J$ and $Q$, respectively. Then, we obtain the computation complexity of Algorithm 1 as $\mathcal{O}(I\log_2(J))$ for $I$ AUVs. Similarly, the computational complexity of Algorithm 2 is $\mathcal{O}(QM)\ast \mathcal{O}(I\log_2(J))$ for $I$ AUVs and $M$ MASSs. For Algorithm 3, to obtain the Bargaining equilibrium of (\textbf{P2}) and (\textbf{P3}) for the tasks of $I$ AUVs, the computational complexity is $\mathcal{O}(I)$.
Therefore, the overall computational complexity of the proposed algorithms is $\mathcal{O}(QM)\ast \mathcal{O}(I\log_2(J))$.}

\section{Performance Evaluation}
In this section, we conduct simulations to validate the effectiveness of our proposed scheme. Specifically, we assess the impact of key parameters on the utility of each party and compare the performance of our proposed method with other benchmark approaches.

\subsection{Simulation setup}
We conduct all the simulations with MATLAB on a PC configured with a Core i7-10510U 1.80 GHz CPU and 8 GB of RAM. 
We consider a satellite-assisted maritime network comprised of one LEOS and five MASSs, which are deployed in a $500\times500\times500$ $m^3$ area, together with 5-30 AUVs capable of autonomous navigation. In the considered scenarion, AUVs make decisions regarding the amount of offloading tasks based on the pricing strategies issued by MASS. We assume AUV $i$ has a total task volume of 50 Mbits. Each AUV communicates with MASS via NOMA, utilizing a channel bandwidth of 1 MHz, and each MASS communicates with LEOS through FDMA, employing a channel bandwidth of 10 MHz. The remaining simulation parameters are listed in Table III.

\begin{table}[htbp]
\centering
\caption{Simulation Parameter Settings}
\begin{tabular}{l|c}%l=left, r=right,c=center分别代表左对齐，右对齐和居中，字母的个数代表列数
\hline
\hline
\quad\quad\quad\quad\quad\quad\quad\quad Parameters &Values  \\ 
\hline
The central frequency of the acoustic signal, $f$\ccc{9377450} &$1$kHz  \\ 
\hline
The ocean noise power, $N_B$\ccc{10100681} &$1\times10^{-4}$dBm  \\
\hline
The air noise power, $N_0$ &$-174$dBm/Hz  \\ 
\hline
Transmission power of AUV $i$, $p_i$ &$0.1$W  \\
\hline
Transmission power of MASS $m$, $p_m$ &$0.5$W  \\
\hline
The coefficient of $d_{i,m}$, $\varrho$\ccc{10118952}&$0.001$  \\
\hline
The elevation angle, $\theta$\ccc{9344666}&$30^{\circ}$  \\
\hline
The computation capacity of AUV $i$, $\rho_i$ &$1\times10^6$ cycles/s  \\
\hline
The power consumption coefficient of AUV $i$, $\varepsilon_i$ &$5\times10^{-20}$  \\
\hline
The computation capacity of MASS $m$, $\rho_n$ &$4\times10^6$ cycles/s\\
\hline
The power consumption coefficient of MASS $m$, $\varepsilon_m$ &$1.5\times10^{-22}$ \\
\hline
The computation capacity of LEOS $n$, $\rho_n$ &$6\times10^8$ cycles/s \\
\hline
The power consumption coefficient of LEOS $n$, $\varepsilon_n$ &$2\times10^{-23}$  \\
\hline
The number of CPU cycles for processing one
bit of \\data in MASS $m$, $\delta_m$\ccc{10100681} &$1\times10^4$ cycles \\
\hline
\hline
\end{tabular}
\end{table}

To validate the effectiveness and efficiency, we further compare the utility performance of the proposed scheme with the following benchmark schemes. 
\begin{itemize}
\item[$\bullet$]
\textcolor{b}{OFDMA based transmission scheme (OTS)\ccc{9765974}: In this scheme, each AUV uploads its workloads to an MASS via OFDMA through underwater acoustic communication, and the communication bandwidth is evenly allocated to each AUV.} 
\item[$\bullet$] 
Offloading Scheme based on Fixed Proportion (OSFP): In this scheme, the offloading ratio $a_{i,m}$  is predetermined and remains constant when AUV $i$ transmits workloads to MASS.
\item[$\bullet$] 
Offloading Scheme based on Random Proportion (OSRP): Similar to\ccc{10197624}, in this scheme, the offloading ratio $a_{i,m}$ is randomly determined when AUV $i$ transmits workloads to MASS.
\item[$\bullet$] 
Offloading Scheme based on Linear Proportion (OSLP): In this scheme, the offloading ratio $a_{i,m}$ varies linearly with the total amount of data when AUV $i$ transmits workloads to MASS. 
\item[$\bullet$]
\textcolor{b}{Hierarchical Adaptive Search Algorithm (HAS)\ccc{9740226}: In this scheme, the optimal solution is progressively attained through iterations. }  
\item[$\bullet$]
Linear Pricing Scheme (LPS): In this scheme, the tender prices of MASS $m$ and LEOS $n$ are linear with the offloading ratio $b_{i,n}$ in the bargaining game\ccc{10100681}.
\item[$\bullet$]
Fixed Pricing Scheme (FPS): In this scheme, the tender prices of MASS $m$ and LEOS $n$ are predetermined and remains constant in the bargaining game.
\item[$\bullet$]
Random Pricing Scheme (RPS): In this scheme, the tender prices of MASS $m$ and LEOS $n$ are randomly determined in the bargaining game\ccc{10449425}. 
\end{itemize}

\subsection{Numerical Results and Analysis}
Fig. 4 illustrates the variation in offloading ratio of AUV $i$ as the number of iterations increases under different \textcolor{b}{link quality}. The proposed scheme consistently converges to a fixed value to achieve the optimal $a_{i,m}$ with the increasing number of iterations. Moreover,  $a_{i,m}$ also increases as the link quality intensifies, indicating that the better the link quality, the more tasks tends to be offloaded by AUV $i$ to conserve its energy consumption. The results presented in Fig. 4 also confirm the efficiency of the proposed BSA algorithm.
\begin{figure}[htbp]
\centerline{\includegraphics[width=0.4\textwidth]{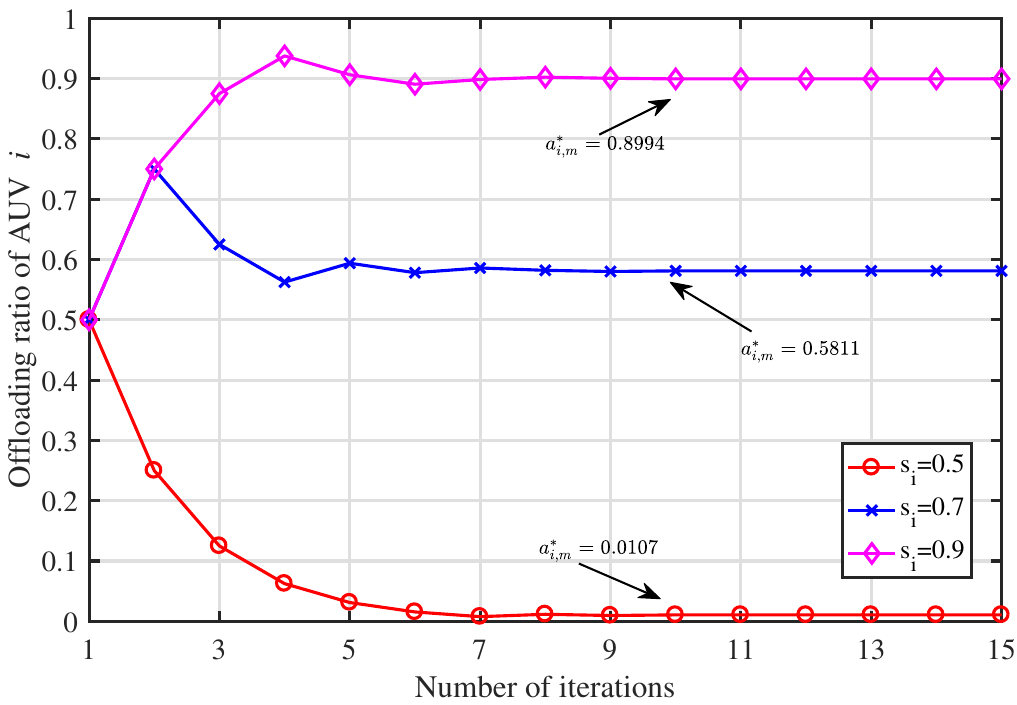}}
\caption{The variation of $a_{i,m}$ under different link quality.}
\label{fig}
\end{figure}

Fig. 5 demonstrates how the utility of AUV $i$ varies with the number of iterations and \textcolor{b}{link quality}. As the number of iterations increases, the algorithm converges to a fixed value for maximum utility. Additionally, the utility function of AUV $i$ escalates as link quality intensifies.
\begin{figure}[htbp]
\centerline{\includegraphics[width=0.4\textwidth]{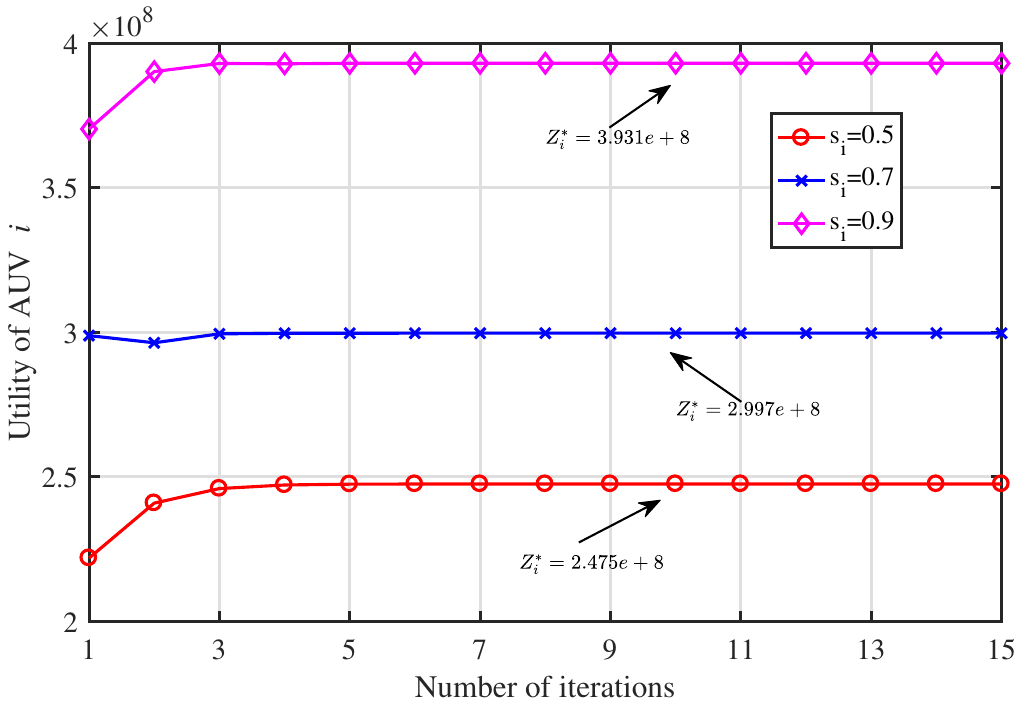}}
\caption{The variation of $Z_i$ under different link quality.}
\label{fig}
\end{figure}

Fig. 6 illustrates the variation in offloading ratio of AUV $i$ as the number of iterations increases under different task volumes. This figure indicates that the proposed algorithm can always converge to a constant value within a finite number of iterations to attain the maximum utility, regardless of how much total workload needs to be processed.
\begin{figure}[htbp]
\centerline{\includegraphics[width=0.4\textwidth]{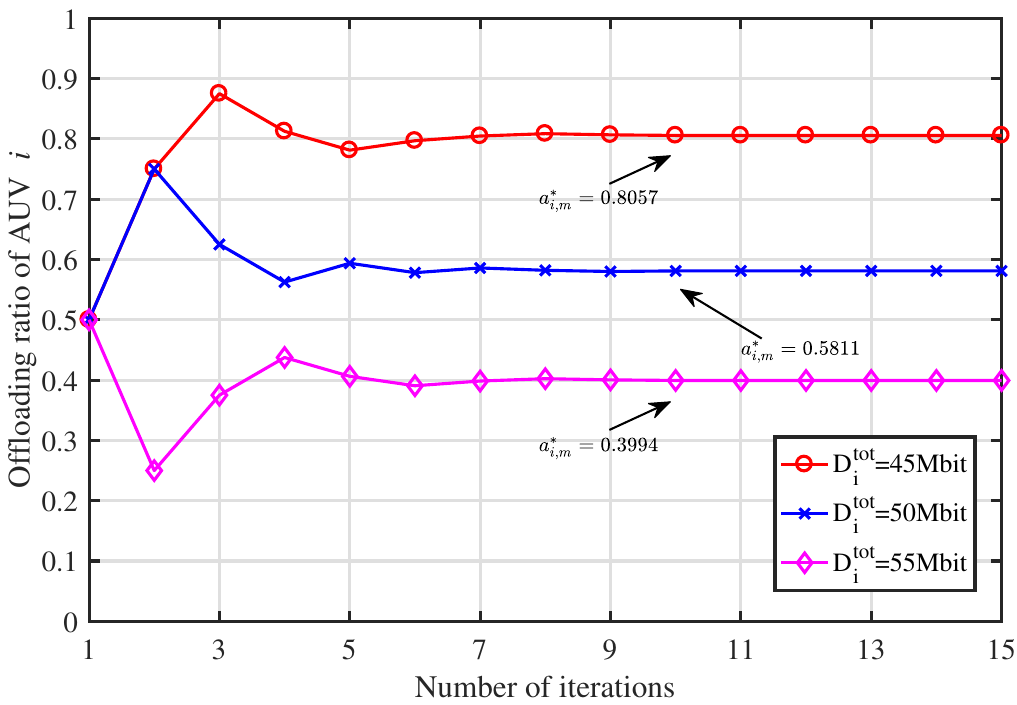}}
\caption{The variation of $a_{i,m}$ under different task volumes.}
\label{fig}
\end{figure}

Fig. 7 depicts the variation in the utility of both MASS and AUV throughout the iterations. As the number of iterations increases, their utilities converge to fixed values to obtain their respective maximums. 
The utility of MASS $m$ increases with higher prices, while the utility of AUV $i$ decreases as prices rise. Nevertheless, once the price surpasses a specific point, both settle into a Stackelberg equilibrium. This figure also verifies the effectiveness of the proposed LSA algorithm.
\begin{figure}[htbp]
\centerline{\includegraphics[width=0.4\textwidth]{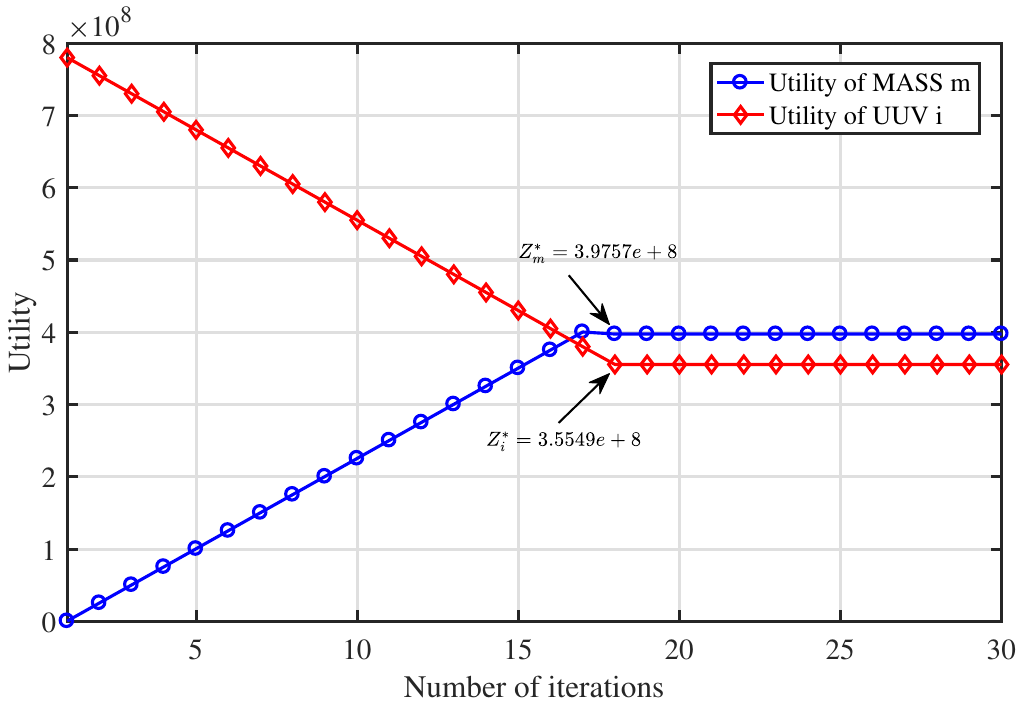}}
\caption{The variation of the utility of MASS $m$ and AUV $i$.}
\label{fig}
\end{figure}

Fig. 8 and Fig. 9 illustrate the comparison of the utilities of MASS and LEOS under varying $b_{i,n}$ and $\eta_i$, respectively. The benchmark is set using different proportion $\eta_i$ for the Bargaining game. As $b_{i,n}$ increases, the utility of MASS decreases, while the utility of LEOS rises, due to the fact that the expenditure paid by MASS escalate with $b_{i,n}$. When $b_{i,n}$ remains constant, an increase in $\eta_i$ leads to a decrease in the utility of MASS and an increase in LEOS's utility, as a larger share of the profits is allocated to LEOS with higher $\eta_i$. The proposed scheme can achieve better utility than others.
\begin{figure}[htbp]
\centerline{\includegraphics[width=0.4\textwidth]{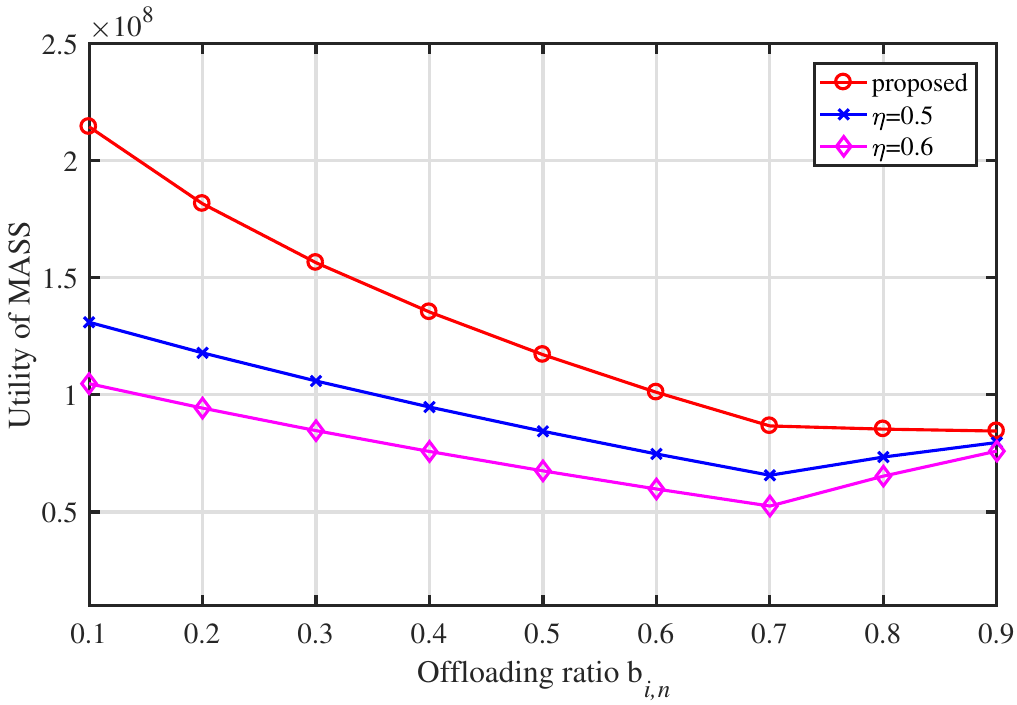}}
\caption{Comparison of the utility of MASS under different $b_{i,n}$.}
\label{fig}
\end{figure}

\begin{figure}[htbp]
\centerline{\includegraphics[width=0.4\textwidth]{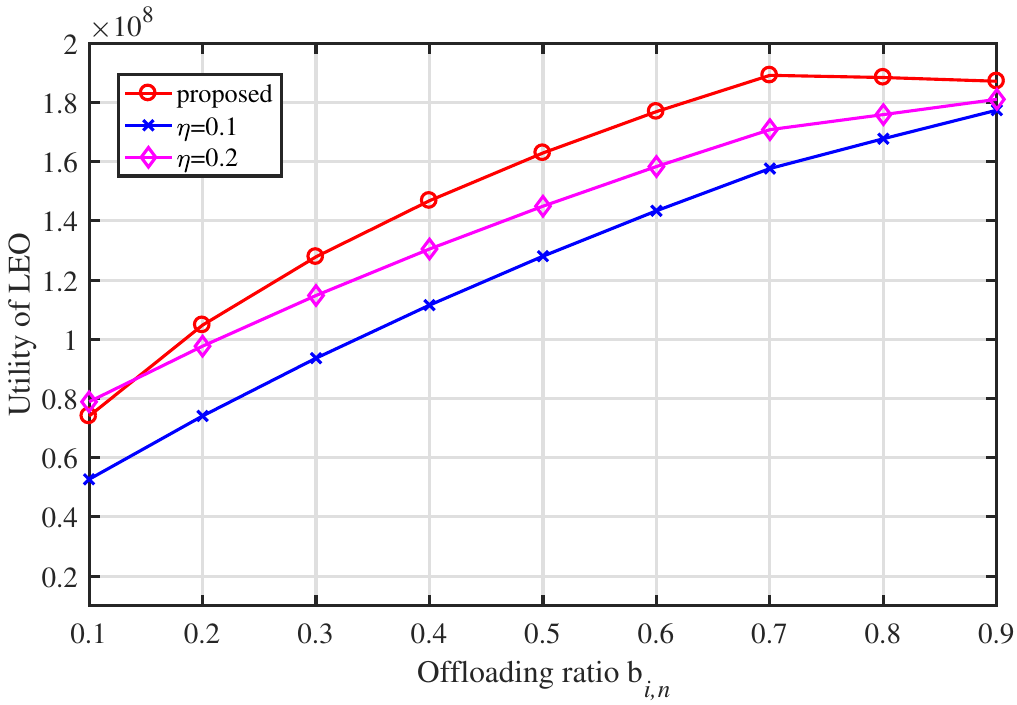}}
\caption{Comparison of the utility of LEOS under different $b_{i,n}$.}
\label{fig}
\end{figure}

Fig. 10 and Fig. 11 demonstrate the utility comparison of the proposed algorithms for AUV and MASS with different number of AUVs. As the number of AUVs escalates, we observe that our proposed scheme has the maximum utility for both the MASS and AUVs, because the proposed scheme takes the optimal offloading ratio $a_{i,m}$ and the optimal pricing strategies of $k_i$ into account. Moreover, as the number of AUVs escalates, the average utility of AUVs rises very slowly due to the non-cooperative nature of the game between AUVs. In contrast, the average utility of MASS increases rapidly as the number of AUVs increases. This is because, with more AUVs, MASS can process more computation tasks, thereby boosting profit margins and enhancing its utility.
\begin{figure}[htbp]\color{b}
\centerline{\includegraphics[width=0.4\textwidth]{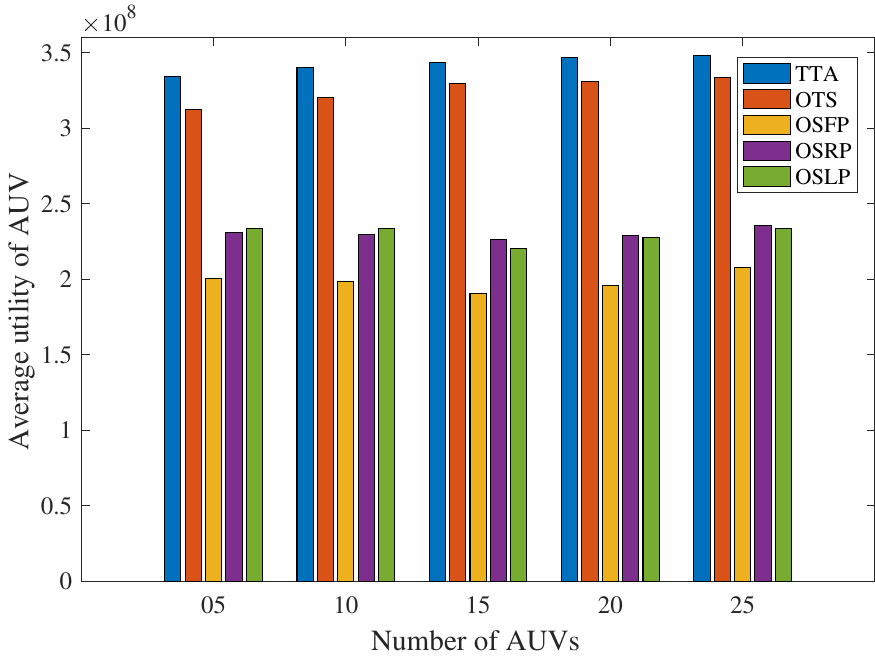}}
\caption{Comparison of the average utility of AUVs with the number of AUVs.}
\label{fig}
\end{figure}

\begin{figure}[htbp]\color{b}
\centerline{\includegraphics[width=0.4\textwidth]{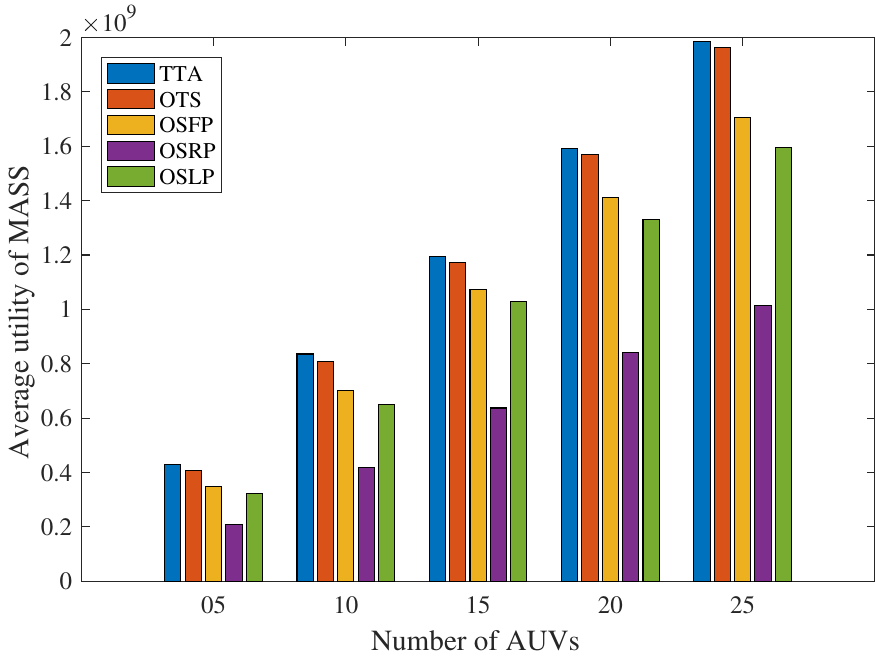}}
\caption{Comparison of the average utility of MASSs with the number of AUVs.}
\label{fig}
\end{figure}

Fig. 12 and Fig. 13 illustrate the utility comparison of our proposed scheme for MASS and LEOS under different values of the offloading ratio $b_{i,n}$. 
With the increment of $b_{i,n}$, we observe that the utility of MASS decreases, while the utility of LEOS increases.
The rationale is that as $b_{i,n}$ increases, a greater proportion of workloads are offloaded to LEOS $n$, prompting MASS $m$ to offer higher expenditure to LEOS $n$. 
Consequently, this enhances the utility of LEOS $n$ while diminishing the utility of MASS $m$. The proposed scheme attains approximate optimal performance due to its consideration of the most advantageous bidding prices for both MASS $m$ and LEOS $n$.
\textcolor{b}{In addition, we see that HAS outperforms the proposed TTA at times in terms of optimization performance, which stems from HAS's ability to progressively converge towards the optimal solution by leveraging global information. However, acquiring global information in real-world applications is challenging, and centralized solutions often entail high time complexity.} 

\begin{figure}[htbp]\color{b}
\centerline{\includegraphics[width=0.4\textwidth]{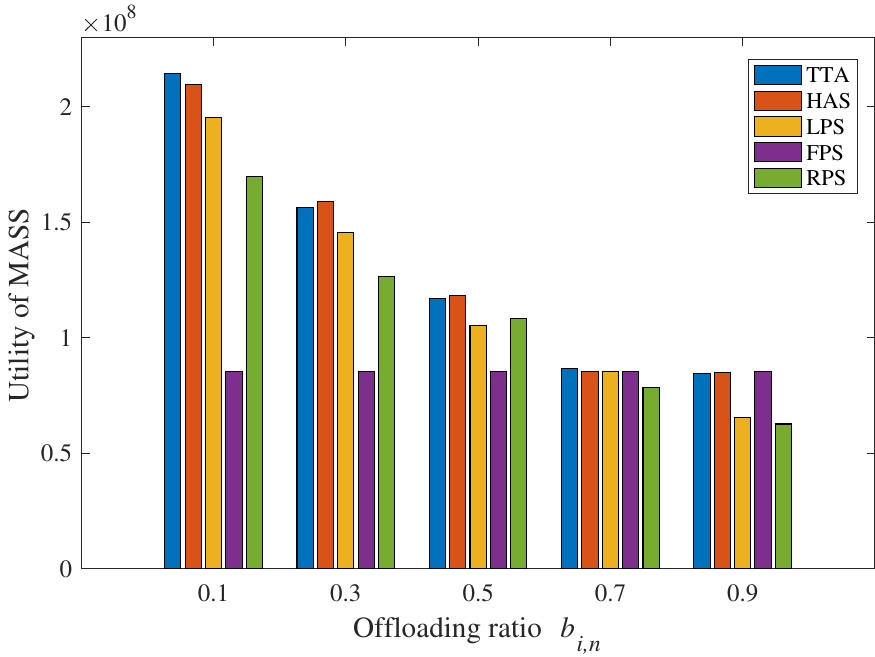}}
\caption{Comparison of the utility of MASSs under different $b_{i,n}$.}
\label{fig}
\end{figure}

\begin{figure}[htbp]\color{b}
\centerline{\includegraphics[width=0.4\textwidth]{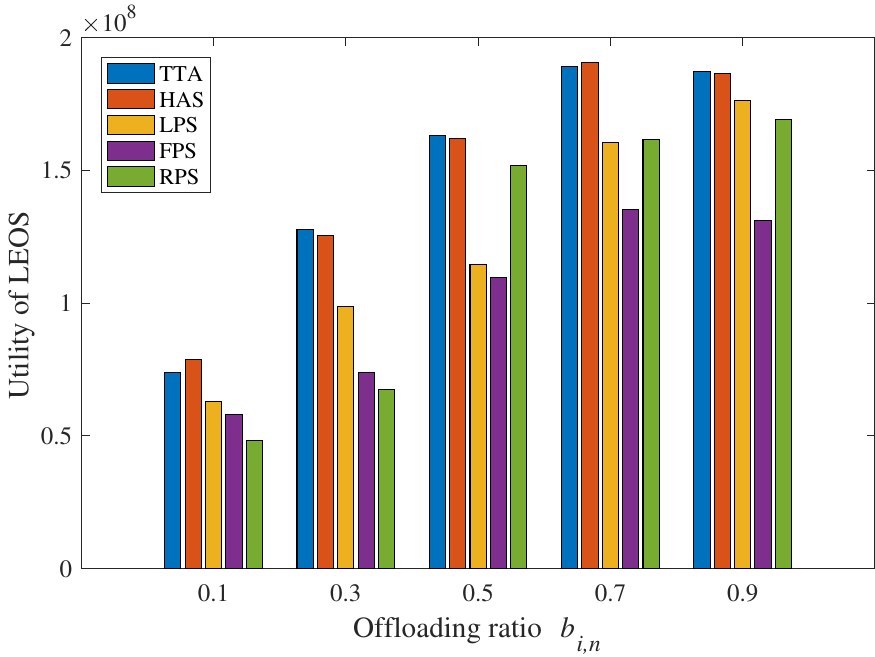}}
\caption{Comparison of the utility of LEOS under different $b_{i,n}$.}
\label{fig}
\end{figure}

\section{Conclusion and Future Work}
In this paper, we have considered a space-sea integrated network architecture and proposed a two-tier task offloading scheme for AUVs using a hybrid Stackelberg-Bargaining game approach to improve the efficiency of task offloading process. Initially, we implement a multi-access task offloading strategy through underwater acoustic communication, enabling AUVs to delegate their workloads to MASSs via NOMA. We formulate the task offloading interactions between AUVs and MASSs as a Stackelberg game to refine the AUV offloading strategy and optimize the pricing strategies of each MASS, to maximize their respective profits. Subsequently, for MASS-to-LEOS transmission, we consider the scenario where each MASS delegates a portion of its workloads to an LEOS via FDMA to avoid the co-channel interference. We formulate the interactions between an MASS and an LEOS as a Bargaining game to optimize their bidding strategies and maximize their respective revenues. Numerical results are presented to validate the efficiency and effectiveness of our proposed scheme. 
%\textcolor{b}{
%Although the proposed two-tier task offloading model offers benefits in terms of maximizing the system efficiency and profitability, it is still subject to the following issues:
%Firstly, the available bandwidth in satellite communication networks is often limited, especially in dense marine environment, which further affects the efficiency of task offloading and the overall system throughput.
%Secondly, the energy consumption of satellite and marine devices during task offloading is another critical limitation. Although task offloading alleviates the computational load on marine devices, the energy expenditure of transmitting data to satellites and receiving responses may be significant, particularly for energy-constrained devices.
%Finally, the variability in marine network conditions, such as channel fading, signal interference, and mobility of marine devices, introduces unpredictability into the task offloading process, making it challenging to maintain a consistent offloading strategy.}
\textcolor{b}{
For future work, we will investigate efficient resource management for satellite-marine integrated networks to improve the overall system performance. In addition, we will explore the development of intelligent algorithms using deep reinforcement learning to improve the adaptability of decision-making process for task offloading in dynamic marine environments.
}

\section*{Acknowledgment}
This work was supported by the National Natural Science Foundation of China under Grant 62371085, Grant 51939001 and Grant 62301411.

\ifCLASSOPTIONcaptionsoff
  \newpage
\fi

% \bibitem{ref1} X. Dou, T. Lyu, B. Fan and D. Liang, ``A Single-Carrier Delay-and-Superposition Based High Spectral Efficiency Differential Chaos Shift Keying for Maritime Internet of Vessels," \emph{IEEE Access}, vol. 10, pp. 128456-128468, Dec. 2022.

% \end{thebibliography}

% \begin{IEEEbiography}[{\includegraphics[width=1.1in,height=1.4in,clip,keepaspectratio]{Xiaoling Han.jpg}}]{Xiaoling Han}
% received the B.S. degree in Electronic Information Engineering from Dalian Maritime University, Dalian, China, in 2020. She is currently working toward the Ph.D. degree at the Department of Information Science and Technology, Dalian Maritime University, Dalian, China, under the supervision of Prof. B. Lin. Her research interests include maritime communication networks, semantic communications and deep learning-based communications.
% 	\end{IEEEbiography}
% 	\vspace{-10 mm} 


%\vfill


\bibliographystyle{ieeetr}
\bibliography{Reference}

\end{document}